\documentclass[preprintnumbers,amsmath,amssymb,prd,floatfix,12pt,superscriptaddress,nofootinbib]{revtex4}
\usepackage{graphicx}
\usepackage{epsfig}
\usepackage{bm}
\usepackage{amsfonts}

\begin{document}

\title{Non-equilibrium thermodynamics of dark energy on the power-law
entropy corrected apparent horizon}
\author{\textbf{M. Umar Farooq}}
\email{mfarooq@camp.nust.edu.pk} \affiliation{Center for Advanced
Mathematics and Physics (CAMP), National University of Sciences and
Technology (NUST), H-12, Islamabad, Pakistan}
\affiliation{Department of Basic Sciences, College of Electrical and
Mechanical Engineering, Peshawar Road, Rawalpindi, Pakistan}

\author{\textbf{Mubasher Jamil}}
\email{mjamil@camp.nust.edu.pk} \affiliation{Center for Advanced
Mathematics and Physics (CAMP), National University of Sciences and
Technology (NUST), H-12, Islamabad, Pakistan}

\begin{abstract}
\textbf{Abstract:} We investigate the Friedmann-Robertson-Walker (FRW)
universe (containing dark energy) as a non-equilibrium (irreversible)
thermodynamical system by considering the power-law correction to the
horizon entropy. By taking power-law entropy area law which appear in
dealing with the entanglement of quantum fields in and out the horizon, we
determine the power-law entropy corrected apparent horizon of the FRW
universe.
\end{abstract}

\maketitle

\newpage

\section{Introduction}

It is quite well known that black hole behaves like a black body, emitting
thermal radiations, with a temperature proportional to its surface gravity
at the black hole horizon and with an entropy which proportional to its
horizon area \cite{beck,hawk}. Further it is also known that the Hawking
temperature and horizon entropy together with the black hole mass obey the
first law of thermodynamics $dM=TdS$ \cite{bard}. In the literature, the
thermodynamical features of black holes have been studied by taking the dark
energy into account \cite{wang1,bousso,wang2,gong} which is equally well
defined in this scenario for the apparent horizon of the black hole, while
the discussion remains no more possible for event horizon.

It is generally believed that the current universe is in accelerated
expansion phase due to the presence of dark energy. The holographic dark
energy (HDE) has been studied extensively in the literature \cite%
{gong1,abda,zhang} and proposed as an interesting candidate of dark energy
which is motivated from the holographic principle \cite{hoof}. In the
derivation of HDE density $\rho _{D}=3c^{2}M_{p}^{2}L^{-2},$ the black hole
entropy $S_{BH}=\frac{A}{4G},$ where $A\backsim L^{2}$ and $A$ represents
the area of the horizon \cite{cohen} plays a key role. Here $M_{p}^{2}=(8\pi
G)^{-1}$ is the modified Planck mass and $3c^{2}$ is constant and introduced
for convenience.

However, this definition of entropy-area relation can be modified due to
power-law corrections to entropy which arise in studying the entanglement of
quantum fields in and out the horizon of the black hole \cite{das}. The
power-law corrected entropy relation is given by the equation \cite{rad}%
\begin{equation}
S=\frac{A}{4G}[1-K_{\alpha }A^{1-\alpha /2}],
\end{equation}%
where $\alpha $ is a dimensionless constant whose value is still matter of
debate, and%
\begin{equation}
K_{\alpha }=\frac{\alpha (4\pi )^{\alpha /2-1}}{(4-\alpha )r_{c}^{2-\alpha }}%
,
\end{equation}%
where $r_{c}$ is the crossover scale. In (1), the second term is defined to
be the power-law correction to the area law mainly appear due to
entanglement, when the wave function is defined in terms of ground state and
excited state \cite{das}. So the entanglement entropy of the ground state
fulfils the Hawking area law while the excited state is related with the
correction, and more excitations produce more deviation from the area law.
For more detail see \cite{das1}. It provides strong basis to entanglement
that it can be used as a\ source of black hole entropy.

It is important to mention that the correction term falls off rapidly with $%
A $ and hence in the semi-classical limit (for large $A$), the area law is
recovered while in the case of small black holes the correction term will
play a significant role. For instance, at low energies, in case of large
horizon area, it is difficult to excite the modes and therefore the ground
state modes contribute to most of the entanglement entropy. On the other
hand, for small area, a large number of field modes will be excited and play
a significant role in the correction causing large deviation from the area
law.

In the cosmological framework, the study of non-equilibrium thermodynamics
of dark energy has some interesting insights. To explore the hidden features
of correction term, Das et al \cite{das} computed leading-order corrections
to the entropy of a thermodynamical system caused by small statistical
fluctuations around equilibrium. They showed that one can obtain a general
logarithmic correction related to the black hole entropy. The
thermodynamical explanation of the interaction between HDE and dark matter
is given in detail in \cite{pav1} and taking the logarithmic correction to
the equilibrium entropy, they derived expression for the interaction term
that was consistent with the observational data. Pavon and Wang \cite{pav2}
proposed a system consists of dark energy and dark matter (with equilibrium
entropies). They argued that if there is transfer of energy between dark
energy and dark matter, the entropy of dark matter directly linked with the
entropy of dark energy. Further, Zhou et al \cite{zho} studied the natural
interaction between dark matter and dark energy in the universe by resorting
to the extended thermodynamics of irreversible process. Moreover, Karami and
Ghaffari \cite{karam} further extended the work of \cite{zho} and examine
the validity of the generalized second-law in non-equilibrium (irreversible)
thermodynamics in a non-flat FRW universe (with dynamical horizon) in which
dark energy interacts with dark matter. In \cite{biao}, Wang and Biao
discussed the non-equilibrium thermodynamics of dark energy on cosmic
apparent horizon. They argued that if the irreversible process is
considered, the proper position for constructing thermodynamics will not be
anymore the apparent horizon. The new position will be related with dark
energy state equation and the irreversible process parameters.

In this paper, our aim is to extend the work done by Wang and Wen-Biao by
introducing the power-law correction to the area-law which appear in dealing
with the entanglement of quantum fields in and out the horizon. Considering
the irreversible process, we find that the non-equilibrium thermodynamic
laws can not be built on the original apparent horizon and develop a
relationship which exhibits how the apparent horizon can be modified to keep
the non-equilibrium thermodynamic law in effect. The outline of the paper is
as follows. Next Section is devoted to the non-equilibrium thermodynamics of
dark energy on the power-law entropy corrected apparent horizon. Concluding
remarks are given in Section III.

\section{Non-equilibrium thermodynamics of dark energy on the power-law
entropy corrected apparent horizon}

The metric of a homogeneous and isotropic universe in FRW model is%
\begin{equation}
ds^{2}=-dt^{2}+a^{2}(t)\left( \frac{dr^{2}}{1-kr^{2}}+r^{2}d\Omega
^{2}\right) ,
\end{equation}%
where $a(t)$ is the scale factor and $k(=-1,0,1)$ is the constant curvature
of their spatial section. By defining $\tilde{r}=a(t)r,$ the location of the
dynamical apparent horizon can be determined by setting $f=0$ in the relation%
\begin{equation}
f=g^{ab}\tilde{r}_{,a}\tilde{r}_{,b}=1-\left( H^{2}+\frac{k}{a^{2}}\right)
\tilde{r}^{2},
\end{equation}%
which is%
\begin{equation}
\tilde{r}=\left( H^{2}+\frac{k}{a^{2}}\right) ^{-\frac{1}{2}}.
\end{equation}%
The observational data suggests that our universe is spatially flat with $%
k=0.$ In this case the apparent horizon is $\tilde{r}=\frac{1}{H}$ which is
equal to the Hubble horizon $R_{A}$ i.e., $R_{A}=\frac{1}{H}.$

We can define the surface gravity $k$ and Hawking temperature respectively\
on the apparent horizon as \cite{gong}%
\begin{equation}
k=-\frac{1}{2}\frac{\partial f}{\partial \tilde{r}}=\frac{\tilde{r}}{%
R_{A}^{2}},
\end{equation}%
and%
\begin{equation}
T=\frac{k}{2\pi }=\frac{\tilde{r}}{2\pi R_{A}^{2}}.
\end{equation}%
So the Hawking temperature at the apparent horizon has the form%
\begin{equation}
T_{A}=T|_{\tilde{r}=R_{A}}=\frac{1}{2\pi R_{A}}.
\end{equation}%
Cai et al \cite{cai} showed that for an apparent horizon of an FRW universe,
there exists Hawking radiations with temperature $\frac{1}{2\pi R_{A}}.$ For
the metric (3), the first Friedmann equation becomes%
\begin{equation}
\left( \frac{\dot{a}}{a}\right) ^{2}=\frac{8}{3}\pi \rho ,
\end{equation}%
where $G=1$ and $\rho $ is the energy density of dark energy. The continuity
equation for dark energy is%
\begin{equation}
\dot{\rho}+3H(1+\omega )\rho =0,
\end{equation}%
where $\omega $ is the parameter of the equation of state of dark energy,
and when $\omega <-\frac{1}{3}$ it exhibits accelerating behavior of the
universe. After defining $\epsilon =\frac{3}{2}(1+\omega ),$ for a definite
equation of state, we obtain%
\begin{equation}
a(t)=t^{\frac{1}{\epsilon }}\text{ \ \ }(0<\epsilon <1)\text{ \ }(-1<\omega
<-\frac{1}{3}),
\end{equation}%
and the apparent horizon can be written as $R_{A}=\frac{1}{H}=\epsilon t.$

The amount of energy flux acrossing the apparent horizon within the time
interval $dt$ is%
\begin{equation}
-dE_{A}=4\pi R_{A}^{2}\rho (1+\omega )dt=\epsilon dt.
\end{equation}%
Since the process of energy flux crossing the apparent horizon is
irreversible, so an internal entropy production will be generated by
this irreversibility. Thus, in the presence of interaction between
dark energy and horizon with in, the time derivative of the
non-equilibrium entropy is expressed as \footnote{In thermodynamics,
the internal entropy is defined as the variation in entropy
\textit{inside} the thermal system, referring to irreversible
processes, defined as $dS=dS_e+dS_i$, where $dS_e$ is the entropy
exchange with the surroundings or across the physical boundary of
the system. Dividing by $dt$ yields
$\frac{dS}{dt}=\frac{dS_e}{dt}+\frac{dS_i}{dt}$ or
$\dot{S}=\dot{S}_{e}+\dot{S}_{i}$. In reversible thermodynamics, the
entropy exchange term is zero and the total entropy is the internal
entropy itself. Note that all natural processes are irreversible
thus the physical universe as a whole is an irreversible thermal
system.}
\begin{equation}
\dot{S}=\dot{S}_{e}+\dot{S}_{i},
\end{equation}%
where $\dot{S}_{i}$ shows the rate of change in internal entropy
production of the universe while $\dot{S}_{e}$ appears as heat flow
between the universe and the horizon. If $\dot{S}_{e}=0$, then this
case corresponds to equilibrium thermodynamics i.e. the study of
thermodynamics dealing with no transfer of energy or entropy across
the physical boundary of the system. We mention here that
irreversible (non-equilibrium) thermodynamics has been studied in
the literature for various dynamical spacetimes in different
gravitational theories \cite{bd}.

Following the steps given in \cite{biao}, we can have%
\begin{eqnarray}
\dot{S}_{i} &=&\oint_{V}\sigma dV, \\
\dot{S}_{e} &=&-\oint\limits_{\Sigma }\vec{J}_{s}.d\vec{\Sigma},
\end{eqnarray}%
where $\sigma $ and $\vec{J}_{s}$ are the internal entropy source\
production density and entropy flow density respectively. Again following
\cite{biao}, by considering the heat conduction between the universe and
horizon, $\sigma $ and $\vec{J}_{s}$ can be expressed as%
\begin{eqnarray}
\sigma  &=&\vec{J}_{q}.\nabla \frac{1}{T}|_{\tilde{r}=R_{A}}, \\
\vec{J}_{s} &=&\frac{\vec{J}_{q}}{T_{A}},
\end{eqnarray}%
where $\vec{J}_{q}$ is the\ heat current. After putting (17) in (15), we
obtain%
\begin{equation}
\dot{S}_{e}=-\frac{1}{T_{A}}\oint\limits_{\Sigma }\vec{J}_{q}.d\vec{\Sigma}=%
\frac{1}{T_{A}}J_{q}A,
\end{equation}%
where we have assumed that $\vec{J}_{q}$ takes the same value at any point
of the apparent horizon with surface $A=4\pi R_{A}^{2}.$ From the power-law
entropy relation (1), we can write%
\begin{equation}
\frac{d}{dt}\left[ \frac{A}{4}(1-K_{\alpha }A^{(1-\frac{\alpha }{2})})\right]
=2\pi R_{A}\dot{R}_{A}\left[ 1-K_{\alpha }\left( \frac{4-\alpha }{2}\right)
(4\pi R_{A}^{2})^{1-\frac{\alpha }{2}}\right] .
\end{equation}%
Taking the relation%
\begin{equation}
d(S)=d_{e}S,
\end{equation}%
we can have%
\begin{equation}
J_{q}=\frac{\epsilon }{4\pi R_{A}^{2}}\left( 1-K_{\alpha }\left( \frac{%
4-\alpha }{2}\right) (4\pi R_{A}^{2})^{1-\frac{\alpha }{2}}\right) ,\text{ \
\ \ }\epsilon =\dot{R}_{A}.
\end{equation}%
According to the Fourier's law%
\begin{equation}
\vec{J}_{q}=-\lambda \nabla T,
\end{equation}%
where $\lambda $ is the thermal conductivity \cite{biao}. So there will be
an energy flux $\vec{J}_{q}$ if there is a $\nabla T$ in a thermodynamical
system. After substituting equations (21) and (22) in (16), we obtain%
\begin{equation}
\sigma =\frac{\epsilon ^{2}}{4\lambda R_{A}^{2}}\left( 1-K_{\alpha }\left(
\frac{4-\alpha }{2}\right) (4\pi R_{A}^{2})^{1-\frac{\alpha }{2}}\right)
^{2}.
\end{equation}%
So after inserting the value of $\sigma $ in (14), we obtain%
\begin{equation}
\dot{S}_{i}=\frac{\epsilon ^{2}\pi R_{A}}{3\lambda }\left( 1-K_{\alpha
}\left( \frac{4-\alpha }{2}\right) (4\pi R_{A}^{2})^{1-\frac{\alpha }{2}%
}\right) ^{2}.
\end{equation}%
Now from (19) and (24) the final expression of irreversible entropy can be
written as%
\begin{eqnarray}
\dot{S} &=&\dot{S}_{e}+\dot{S}_{i}  \nonumber \\
&=&2\pi R_{A}\epsilon \left( 1-K_{\alpha }\left( \frac{4-\alpha }{2}\right)
(4\pi R_{A}^{2})^{1-\frac{\alpha }{2}}\right) \left( 1+\frac{\epsilon }{%
6\lambda }-\frac{\epsilon }{6\lambda }K_{\alpha }\left( \frac{4-\alpha }{2}%
\right) (4\pi R_{A}^{2})^{1-\frac{\alpha }{2}}\right) .
\end{eqnarray}%
Therefore, the first law in irreversible thermodynamics holds if we define%
\begin{equation}
dS=-\frac{dE}{\tilde{T}_{A}}=2\pi \tilde{R}_{A}\epsilon dt,
\end{equation}%
where $\tilde{T}_{A}=\frac{1}{2\pi \tilde{R}_{A}}$ and%
\begin{equation}
\tilde{R}_{A}=R_{A}\left[ (1-K_{\alpha }(\frac{4-\alpha }{2})(4\pi R_{A})^{1-%
\frac{\alpha }{2}})\left( 1+\frac{\epsilon }{6\lambda }-\frac{\epsilon }{%
6\lambda }K_{\alpha }\left( \frac{4-\alpha }{2}\right) (4\pi R_{A}^{2})^{1-%
\frac{\alpha }{2}}\right) \right] ,
\end{equation}%
which is also called power-law entropy corrected apparent horizon. To check
consistency, if we put $\alpha =0,$ we obtain%
\begin{equation}
\tilde{R}_{A}=R_{A}\left( 1+\frac{\epsilon }{6\lambda }\right) ,
\end{equation}%
which is the same as worked out in \cite{biao}.

\section{Conclusion}

In this paper, we considered the FRW universe and investigated
non-equilibrium feature of it by taking the power-law correction to
the horizon entropy. In an irreversible scenario, as energy goes
outside the horizon, an internal entropy production term plays a
significant role, which comes out to zero in the case of
equilibrium. In \cite{biao}, they showed that by considering the
irreversible process, the non-equilibrium thermodynamic laws can not
hold true on original apparent horizon and should be modified. They
derived the expression for the modified apparent horizon which
depends on the state parameter of dark energy $\epsilon $ and a
non-equilibrium factor $\lambda .$ Here we extended their study by
taking the power-law correction to entropy which appear in dealing
with the entanglement of quantum fields in and out the horizon. The
appearance of power-law correction terms to the horizon entropy is a
fundamental prediction of quantum entanglement and must be
considered in performing the thermodynamics study of horizon for the
FRW universe. So from the power-law correction to entropy, we
determined the entropy corrected form of the apparent horizon of the
FRW universe.

Since the power-law entropy corrected apparent horizon is a
cosmological horizon, hence it can be useful in variety of ways. For
instance, in $f(T)$ gravity, the thermodynamic properties of this
horizon can be investigated following \cite{bamba}. Also in
fractional action cosmology, one can study the generalized second
law of thermodynamics at the the power-law entropy corrected
apparent horizon following \cite{ujjal}. Similarly one can
investigate thermal properties of interacting holographic and
new-agegraphic dark energy models with corrected-apparent horizon
\cite{j}.

\subsubsection*{Acknowledgment}
We would like to thank the referees for giving useful comments on
our work.

\end{document}